\documentclass[preprint]{aastex}

\usepackage{natbib}
\newcommand{\beq}{\begin{equation}}
\newcommand{\beqa}{\begin{eqnarray}}
\newcommand{\eeq}{\end{equation}}
\newcommand{\eeqa}{\end{eqnarray}}

\newcommand{\simgt}{\lower.5ex\hbox{$\; \buildrel > \over \sim \;$}}
\newcommand{\simlt}{\lower.5ex\hbox{$\; \buildrel < \over \sim \;$}}

\newcommand{\psim}{\mbox{\raisebox{-1.0ex}{$\stackrel{\textstyle >}
{\textstyle \sim}$ }}}
\newcommand{\bd}[1]{\mbox{\boldmath $#1$}}

\slugcomment{To appear in ApJ}

\shorttitle{Lensing Minkowski Functionals}
\shortauthors{Shirasaki et al.}

\begin{document}

\title{
Probing Primordial Non-Gaussianity with Weak Lensing Minkowski Functionals
}

\author{
  Masato Shirasaki, 
  Naoki Yoshida
}
\affil{Kavli Institute for the Physics and Mathematics of the Universe,
University of Tokyo, Kashiwa, Chiba 277-8583, Japan\\
Department of Physics, University of Tokyo, Tokyo 113-0033, Japan
}
\email{masato.shirasaki@ipmu.jp}

\author{Takashi Hamana}
\affil{National Astronomical Observatory of Japan, Tokyo 181-0015, Japan}

\and

\author{Takahiro Nishimichi}
\affil{Kavli Institute for the Physics and Mathematics of the Universe,
University of Tokyo, Kashiwa, Chiba 277-8583, Japan
}

\begin{abstract}
We study the cosmological information contained in the 
Minkowski Functionals (MFs) of weak gravitational lensing convergence maps.
We show that the MFs provide strong constraints on the local type primordial 
non-Gaussianity parameter $f_{{\rm NL}}$.
We run a set of cosmological $N$-body simulations and perform 
ray-tracing simulations of weak lensing,
to generate 100 independent convergence maps of 25 deg$^{2}$ field-of-view
for $f_{\rm NL} = -100, 0$ and $100$.
We perform a Fisher analysis to study the degeneracy among other cosmological 
parameters such as the dark energy equation of state parameter $w$ 
and the fluctuation amplitude $\sigma_8$. We use fully nonlinear covariance
matrices evaluated from 1000 ray-tracing simulations.
For the upcoming wide-field observations such as Subaru Hyper Suprime-Cam survey 
with the proposed survey area of 1500 deg$^{2}$, 
the primordial non-Gaussianity can be constrained 
with a level of $f_{{\rm NL}} \sim 80$ and $w \sim 0.036$
by weak lensing MFs.
If simply scaled by the effective survey area, 
a 20000 deg$^{2}$ lensing survey using Large Synoptic Survey Telescope 
will give constraints of $f_{{\rm NL}} \sim 25$ and $w \sim 0.013$.
We show that these constraints can be further improved by a tomographic 
method using source galaxies in multiple redshift bins. 
\end{abstract}


\keywords{Cosmology: cosmological parameters --- large-scale structure of Universe}

\section{INTRODUCTION}

An array of recent precise cosmological observations such as
the cosmic microwave background (CMB) anisotropies \citep{2011ApJS..192...18K} 
and the large-scale structure \citep[e.g.][]{2006PhRvD..74l3507T,2010MNRAS.404...60R}
established the so-called standard cosmological model,
in which the energy content of the present-day universe is
dominated by dark energy and dark matter.
The standard model assumes that the primordial density fluctuations are
generated via inflation in the very early universe, which
seeded eventually all the rich structures of the universe we see today.
The density fluctuations generated through the mechanism generally
follow Gaussian statistics. Deviations from primordial Gaussianity 
would provide interesting information on the early universe.
For example, some inflationary models predict generation of 
non-Gaussian density fluctuations, 
called the primordial non-Gaussianity 
\citep[for a review, see][]{2004PhR...402..103B}.

For the so-called local type non-Gaussian models,
the initial curvature fluctuation
$\Phi$ is expressed by the Taylor expansion of a Gaussian field $\Phi_{\rm G}$
\beqa
\Phi(\bd{x}) = \Phi_{\rm G}(\bd{x}) + f_{{\rm NL}} \left( \Phi_{\rm G}^2(\bd{x}) - \langle \Phi_{G}^2(\bd{x}) \rangle \right) + \cdots. \label{eq:fNL}
\eeqa
The leading coefficient $f_{\rm NL}$ determines the strength of the non-Gaussianity
\citep{2001PhRvD..63f3002K}.
The amplitude of $f_{\rm NL}$ depends on the perturbation generation mechanism and
hence on the physics of inflation.

There have been several observational probes of $f_{\rm NL}$.
For example, the bispectrum of the CMB anisotropies is shown to be a powerful
probe of $f_{\rm NL} $ \citep{2011ApJS..192...18K}.
The abundance and the distribution of galaxies and galaxy 
clusters can be also used to constrain $f_{\rm NL}$.
Primordial non-Gaussianity induces a strong scale-dependence in  
the power and bi-spectra of biased objects at large length scales 
and changes the abundance of very massive clusters 
\citep[e.g.][]{2008PhRvD..77l3514D,2010JCAP...07..002N,2011JCAP...03..017S}.
It is important to note that the large-scale clustering 
of galaxies and galaxy cluster 
can be used to constrain or possibly detect 
scale-dependent non-Gaussianity which evades the CMB constraints
\citep[e.g.][]{2008JCAP...04..014L}.
\citet{2010PhRvD..82h3511D} derived constraints on $f_{\rm NL}$ 
by combining the CMB anisotropies and galaxy clustering data.
However, galaxies are thought to be biased tracers
of underlying matter distribution.
\cite[e.g.][]{2007MNRAS.374.1303C, 2010JCAP...07..013R}.
In order to avoid uncertainties associated with complicated
galaxy bias, it would be ideal probe the dark matter distribution directly.

Gravitational lensing is a powerful method to study dark matter 
distribution \citep[e.g.][]{2012MNRAS.420.3213O}.
Future weak lensing surveys are aimed at measuring cosmic shear
over a wide area of more than a thousand square degrees. 
Such observational programmes include 
Subaru Hyper Suprime-Cam (HSC) \footnotemark[1],  
the Dark Energy Survey (DES) \footnotemark[2], 
and the Large Synoptic Survey Telescope (LSST) \footnotemark[3].
\footnotetext[1]{\rm{http://www.naoj.org/Projects/HSC/j\_index.html}} 
\footnotetext[2]{\rm{http://www.darkenergysurvey.org/}}
\footnotetext[3]{\rm{http://www.lsst.org/lsst/}}
Space missions such as Euclid and WFIRST are also promising 
to conduct a very wide-field cosmology survey.
The large set of cosmic shear data will enable us to greatly improve 
the constraints on cosmological parameters which include 
primordial non-Gaussianity \citep[e.g.][]{2011PhRvD..83b3008O}.

There are many statistics proposed to characterize the
large-scale matter distribution.
Minkowski Functionals (MFs) are among useful statistics to extract 
the non-Gaussian information from two-dimensional or three-dimensional maps.
For example, the full set of CMB MFs has already given comparable constraints 
to those obtained using the CMB bispectrum 
\citep{2008MNRAS.389.1439H}.
\citet{2001ApJ...552L..89M} and \citet{2001ApJ...551L...5S}
studied $\Omega_{m}$-dependence of weak lensing MFs. 
More recently, \citet{2012PhRvD..85j3513K} showed that the lensing 
MFs contain significant cosmological information, beyond the
power-spectrum.
It is important and timely to study weak lensing MFs using fully nonlinear
simulations of cosmic structure formation.

In this paper, we forecast for future weak lensing surveys
using MFs. In particular, we examine their ability  
to constrain the amplitude of the local-type primordial 
non-Gaussianity.
We run a large set of $N$-body simulations and then
perform ray-tracing simulations of gravitational lensing.
We measure MFs directly from the highly-resolved mock cosmic shear maps.
We perform a full Fisher analysis to study the degeneracy among other cosmological 
parameters, especially the dark energy equation of state $w=P/\rho$.
We propose to use future cosmology surveys to constrain, or possibly
detect, primordial non-Gaussianities.

The rest of the present paper is organized as follows.
In Section \ref{sec:sim}, we describe the details of $N$-body simulations
and our ray-tracing simulations of gravitational lensing.
In Section \ref{sec:MFs}, we summarize the basics of MFs.
In Section \ref{sec:res}, we show the results of a Fisher analysis using MFs. 
We clarify the degeneracy among the three parameters we consider.
Concluding remarks and discussions are given in Section~\ref{sec:conc}.

\section{METHODOLOGY}
\label{sec:sim}
\subsection{$N$-body Simulations}
\label{sim}
We run a number of cosmological $N$-body simulations to generate
weak lensing convergence maps. 
We use the parallel Tree-Particle Mesh code
 {\tt Gadget2} \citep{2005MNRAS.364.1105S}.
Each simulation is run with $256^3$ dark matter particles.
We run simulations of two different volumes, $240$ and $480 \ h^{-1}$Mpc
on a side. 
We generate the initial conditions following a parallel code 
developed in \citet{2009PASJ...61..321N} and
\citet{2011A&A...527A..87V}, which employs the 
second-order Lagrangian perturbation theory 
\cite[e.g.][]{2006MNRAS.373..369C}.
The initial redshift is set to 
 $z_{\rm init}=50$, where we compute the linear matter transfer function using
 {\tt CAMB} \citep{Lewis:1999bs}.
We then follow \citet{2010JCAP...07..002N} 
to add non-Gaussian corrections to the initial conditions. 
For our fiducial cosmology, we adopt the following parameters:
matter density $\Omega_{m}=0.2726$, dark energy density $\Omega_{\Lambda}=0.7274$, 
Hubble parameter $h=0.704$ and 
the scalar spectral index $n_s=0.963$.
These parameters are consistent with 
the WMAP 7-year results \citep{2011ApJS..192...18K}.
For the primordial non-Gaussianity parameter, 
we adopt $f_{\rm NL}=0, \pm100$.
To investigate the degeneracy of the cosmological parameters, 
we also run the same set of simulations for different $w$ and $\sigma_{8}$,
where $\sigma_{8}$ is the {\it rms} of the density field on 8 $h^{-1}$Mpc. 
For these runs, we fix the amplitude of curvature fluctuations 
$\Delta_{\cal R}^{2}=2.441\times 10^{-9}$ 
at the pivot scale $k=0.002{\rm Mpc}^{-1}$.
For $w=-0.8, -1.0,$ and $-1.2$, the resulting 
$\sigma_{8}$ is equal to 0.753, 0.809, and 0.848, respectively.
We summarize the simulation parameters in Table \ref{tab:nbody}.

 After performing the simulations with $w=-0.8$ and $w=-1.2$,
we found that our code had a bug in converting the physical time
to the cosmic expansion parameter in the part where gravitational 
acceleration due to particle-particle interactions is calculated. 
This bug affected the results for $w=-0.8$ and $w=-1.2$. However, 
we have explicitly checked that the effect was very minor and that
the statistics we use below were hardly affected. For example, 
the matter power spectra at $z=0$ after and before the bug was
fixed differ less than 0.1 percent in the power amplitude
at $0.04 < k < 1 \, h/{\rm Mpc}$. 


\subsection{Ray Tracing Simulations}

For ray-tracing simulations of gravitational lensing, 
we generate light-cone outputs using multiple simulation boxes
in the following manner. The small- and large-volume simulations are 
placed to cover a past light-cone of a hypothetical observer 
with angular extent $5^{\circ}\times 5^{\circ}$, from redshift 
$z=0$ to $z\sim 3.5$, similarly to the methods in  
\citet{2000ApJ...537....1W}
and 
\citet{2001MNRAS.327..169H}.
We follow
\citet{2009ApJ...701..945S}
in order to simulate gravitational lensing signals.
Details of the configuration are found there.

We set the initial ray directions on $2048^2$ grids.
The corresponding angular grid size is 
$5^{\circ}/2048\sim 0.15$ arcmin.
To avoid the same structure aligned along the line of sight, 
we shift randomly the $N$-body simulation boxes.
In addition, we use simulation outputs from independent
realizations when generating the light-cone outputs.
We generate 100 independent convergence maps 
from 20 $N$-body simulations
for each cosmological model.
We fix the redshift of the source 
galaxies to $z_{\rm source} = 1.0, 1.5$.

It is well-known that the intrinsic 
ellipticities of source galaxies induce noises to lensing shear maps.
Assuming intrinsic ellipticities are uncorrelated,
we compute the noise to convergence as
\begin{eqnarray}
\langle \kappa_{\rm noise}(x,y)\kappa_{\rm noise}(x^{\prime}, y^{\prime})\rangle
 = \frac{\sigma_{\gamma}^{2}}{n_{\rm gal}A_{\rm pix}}\delta_{xx^{\prime}}\delta_{yy^{\prime}},
\label{eq:noise}
\end{eqnarray}
where $\delta_{xx^{\prime}}$ is the Kronecker delta symbol, 
$n_{\rm gal}$ is the number density of source galaxies,
$A_{\rm pix}$ is the solid angle of a pixel, and
$\sigma_{\gamma}$ is the {\it rms} of the shear noise.
Throughout this paper, we adopt $\sigma_{\gamma}=0.4$ and 
assume that the total number density of the source galaxies is 30 galaxies/${\rm arcmin}^{2}$.
These are typical values for a weak lensing survey using Subaru telescope 
\citep[e.g.][]{2007ApJ...669..714M}. 
When we study a tomographic method (see Section \ref{sec:conc}), 
we assume $n_{\rm gal}=15 \ {\rm galaxies}/{\rm arcmin}^{2}$ at $z_{\rm source}=1.0$ and $1.5$
\footnote{
With this simple split, the {\it total} noise per pixel can be kept constant. 
We can then study the significance of the high-z source galaxies for parameter constraints.
}.
To simulate a more realistic survey, we add the Gaussian noises following 
Eq.(\ref{eq:noise}) to our simulated maps. 
Then we perform the Gaussian filtering 
to the noisy lensing maps.
We set the smoothing scale to 1 arcmin.
This choice corresponds to the optimal 
smoothing scale for the detection of massive halos 
using weak lensing with $z_{\rm source} = 1.0$
\citep{2004MNRAS.350..893H}.
We discuss the effect of smoothing on the statistical analysis 
in Section \ref{sec:conc}.

\section{MINKOWSKI FUNCTIONALS}
\label{sec:MFs}
\subsection{Basics}

Minkowski Functionals are morphological statistics for some smoothed random field 
above a certain threshold.
In general, for a given $D$-dimensional smoothed field, 
one can calculate $D+1$ MFs $V_{i}$.
On $\mathbb{S}^2$, one can define 2+1 MFs $V_{0}, V_{1}$ and $V_{2}$.
$V_{0}$, $V_{1}$ and $V_{2}$ describe the fraction of area above the threshold, the total boundary length of contours, 
and the integral of the geodesic curvature $K$ along the contours.
Mathematically, for a given threshold $\nu$, MFs are defined as
\begin{eqnarray}
V_{0}(\nu) &\equiv& \frac{1}{4\pi}\int_{Q_{\nu}}\, {\rm d}S, \label{eq:V0_def} \\
V_{1}(\nu) &\equiv& \frac{1}{4\pi} \int_{\partial Q_{\nu}}\, \frac{1}{4} {\rm d}\ell , \label{eq:V1_def} \\
V_{2}(\nu) &\equiv& \frac{1}{4\pi} \int_{\partial Q_{\nu}}\, \frac{1}{2\pi}K{\rm d}\ell , \label{eq:V2_def}
\end{eqnarray}
where $Q_{\nu}$ and $\partial Q_{\nu}$ represent the excursion set 
and the boundary of the excursion set for a smoothed field $u(\bd{\theta})$.
They are given by
\beqa
Q_{\nu} = \{\bd{\theta}\, |\, u(\bd{\theta}) > \nu\}, \\
\partial Q_{\nu} =\{ \bd{\theta} \, |\, u(\bd{\theta}) = \nu \}.
\eeqa

\begin{figure}
\begin{center}
        \includegraphics[width=0.3\columnwidth]{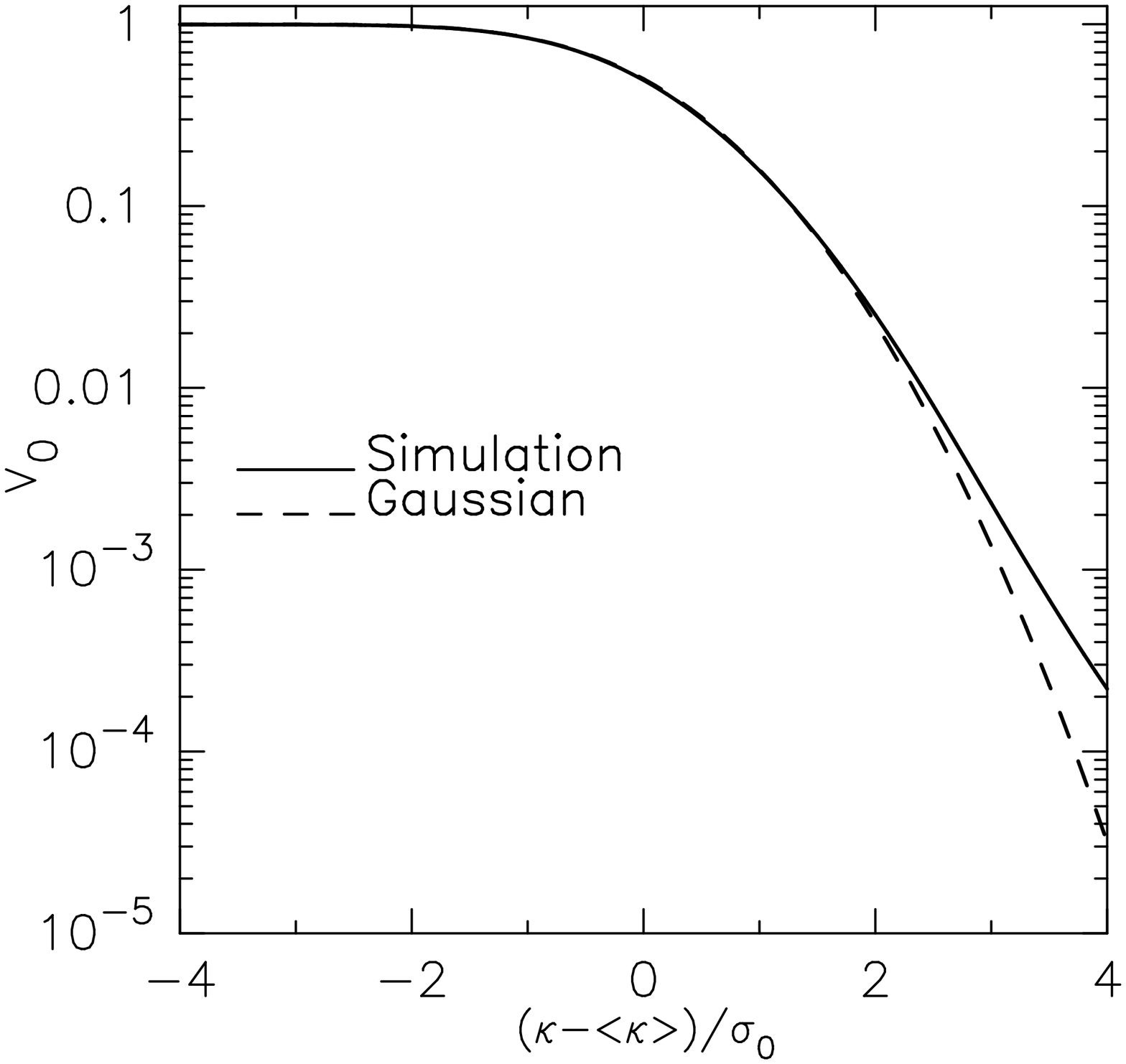}
        \includegraphics[width=0.3\columnwidth]{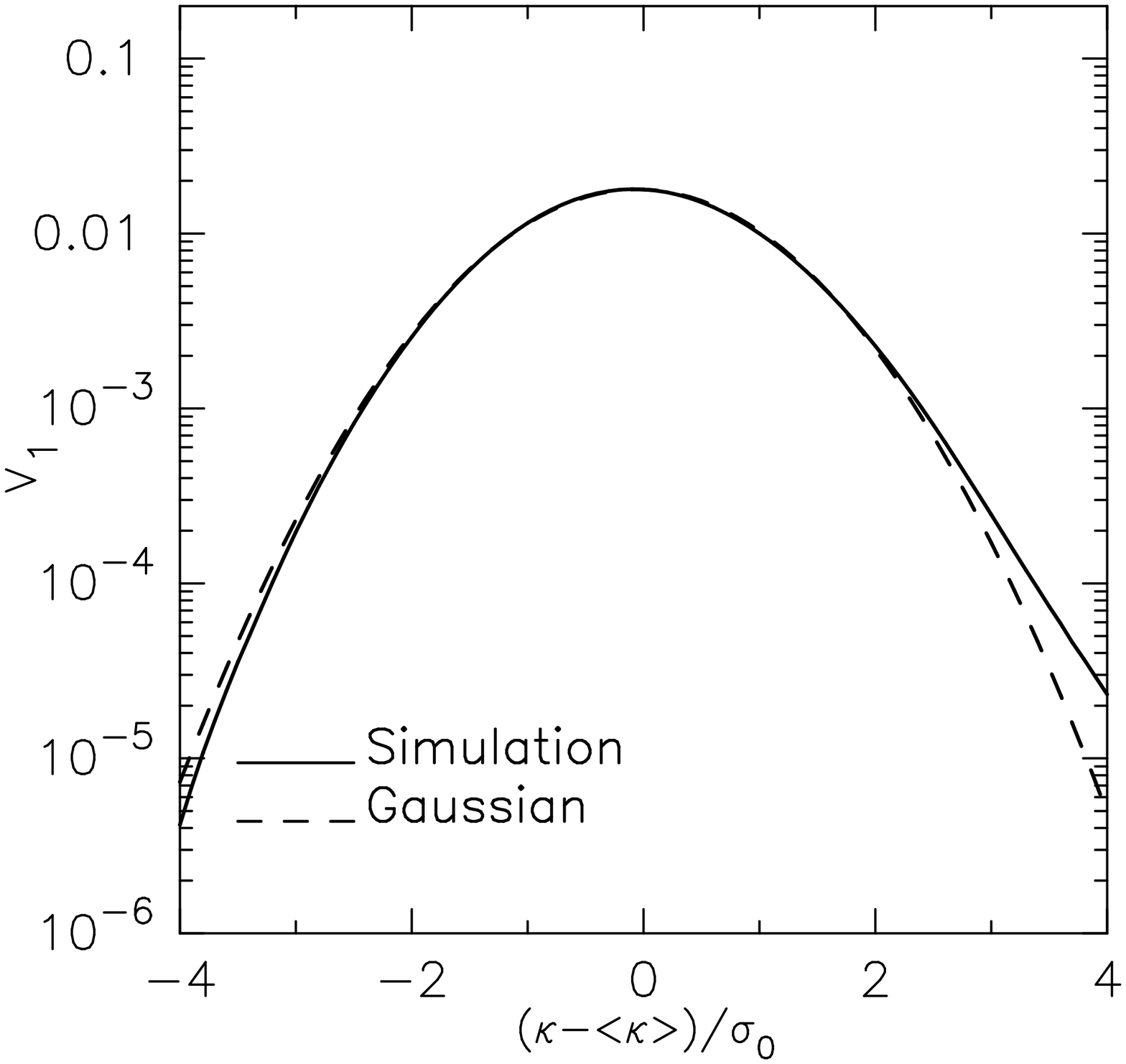}
        \includegraphics[width=0.3\columnwidth]{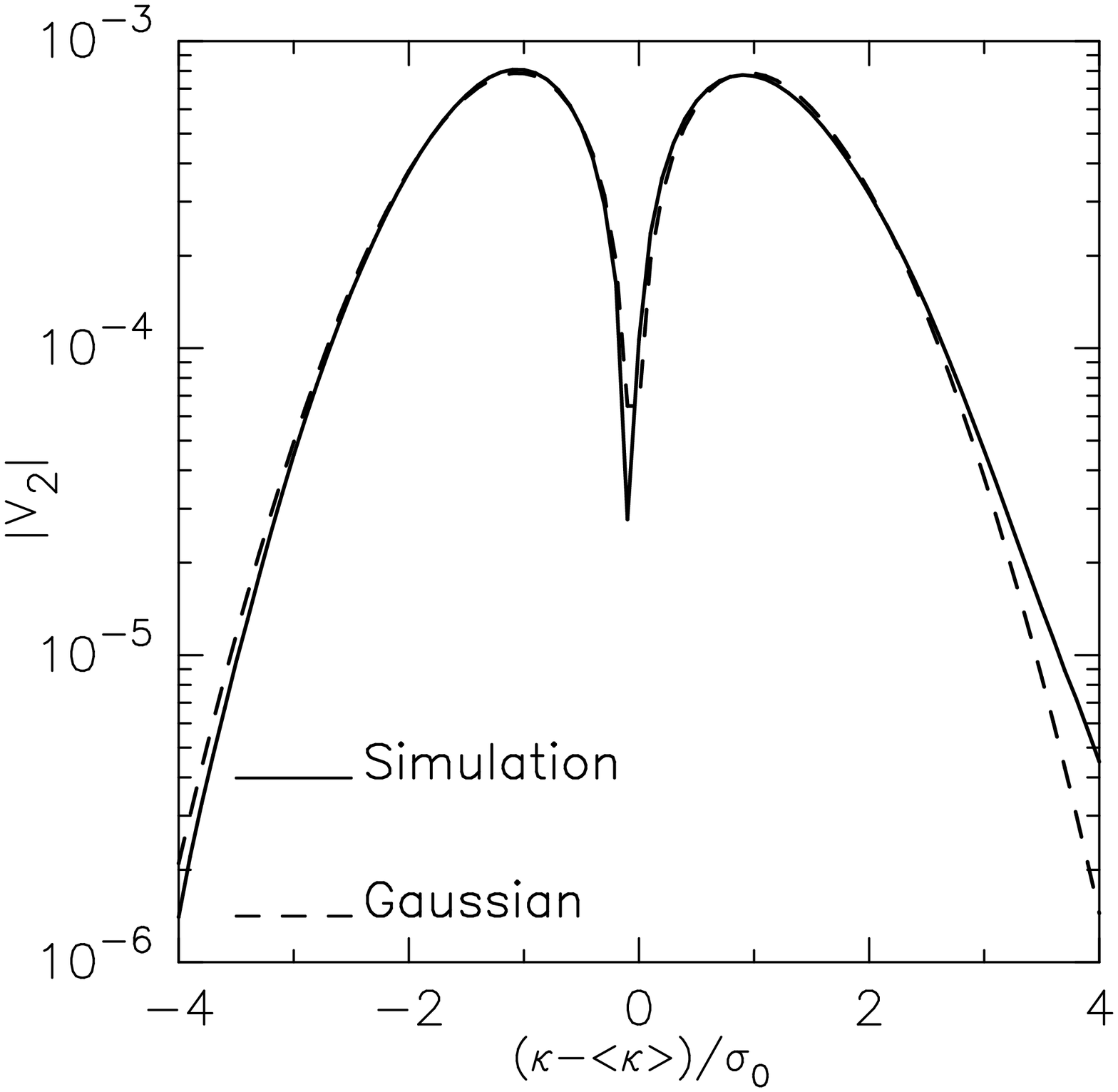}
	\caption{
	The Minkowski Functionals from our simulated convergence maps.
	The solid lines show the average MFs over 100 maps with the fiducial cosmological parameters.
	The dashed lines show the analytic formula of MFs for the Gaussian field \citep{1986PThPh..76..952T}.
	We use 100 convergence maps located on $z_{\rm source} = 1$.
	\label{fig:MFs_ex}
	}
    \end{center}
\end{figure}

We follow \citet{2012JCAP...01..048L}
to calculate the MFs from pixelated convergence maps.
In this step, we convert a 
convergence field $\kappa$ into $x = (\kappa - \langle \kappa \rangle)/\sigma_{0}$
where $\sigma_{0}$ is the standard deviation of a noisy convergence field 
on a $5^{\circ} \times 5^{\circ}$ map.
In binning the thresholds, we set $\Delta x = 0.1$ from $x=-5$ to $x = 5$.
We have checked that the binning is sufficient to 
reproduce the analytic MFs formula \citep{1986PThPh..76..952T} 
for mock 1000 maps of Gaussian random fields.
Figure \ref{fig:MFs_ex} shows the measured and averaged MFs for our 100 convergence maps 
for the $\Lambda$CDM model and for $z_{\rm source} = 1$.
We also plot the analytic formula of MFs for Gaussian statistics to show the non-Gaussian
features of the simulated convergence maps.

\subsection{Dependence on $f_{\rm NL}$}

\begin{figure}
\begin{center}
        \includegraphics[clip, width=0.3\columnwidth]{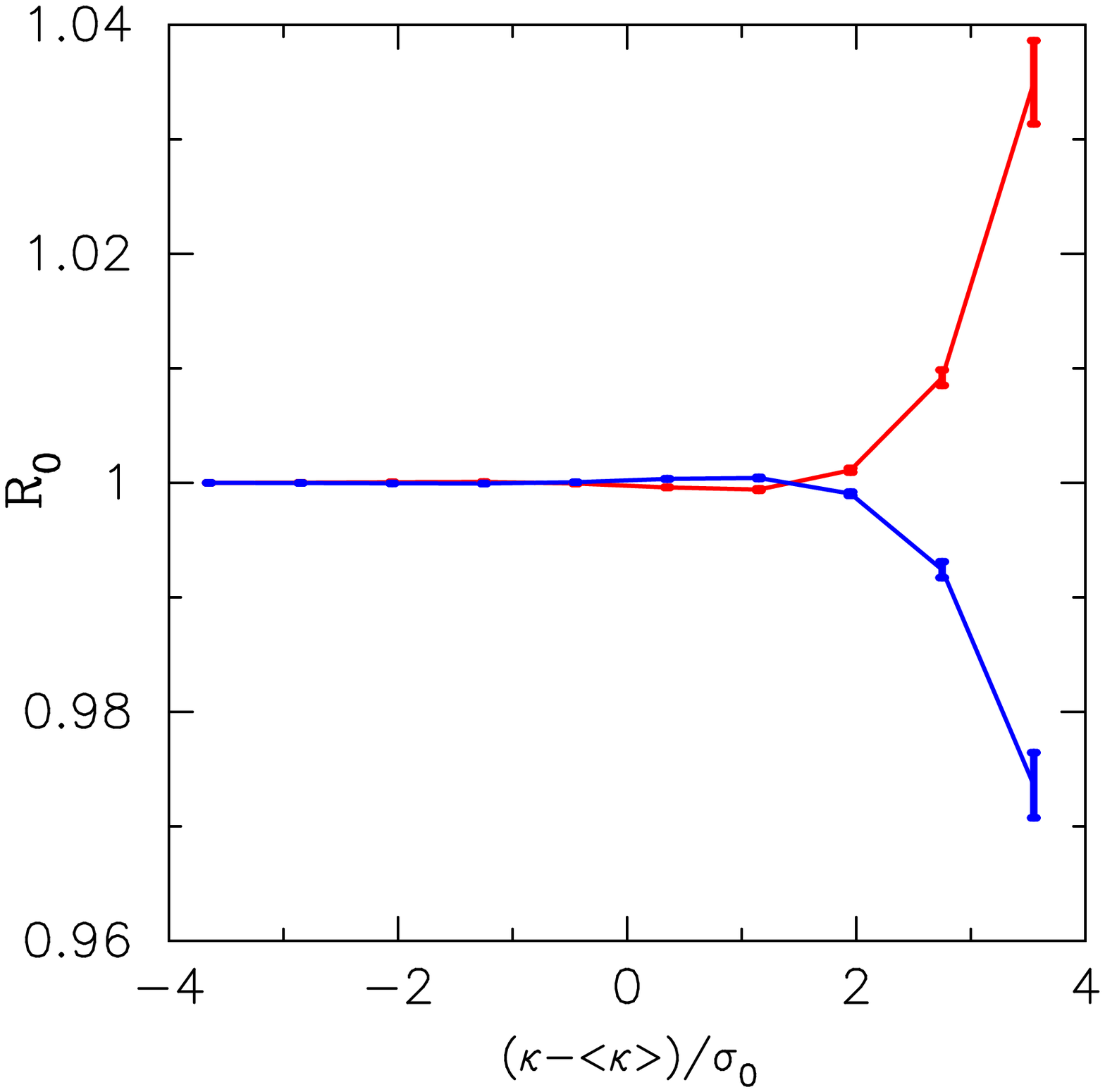}
        \includegraphics[clip, width=0.3\columnwidth]{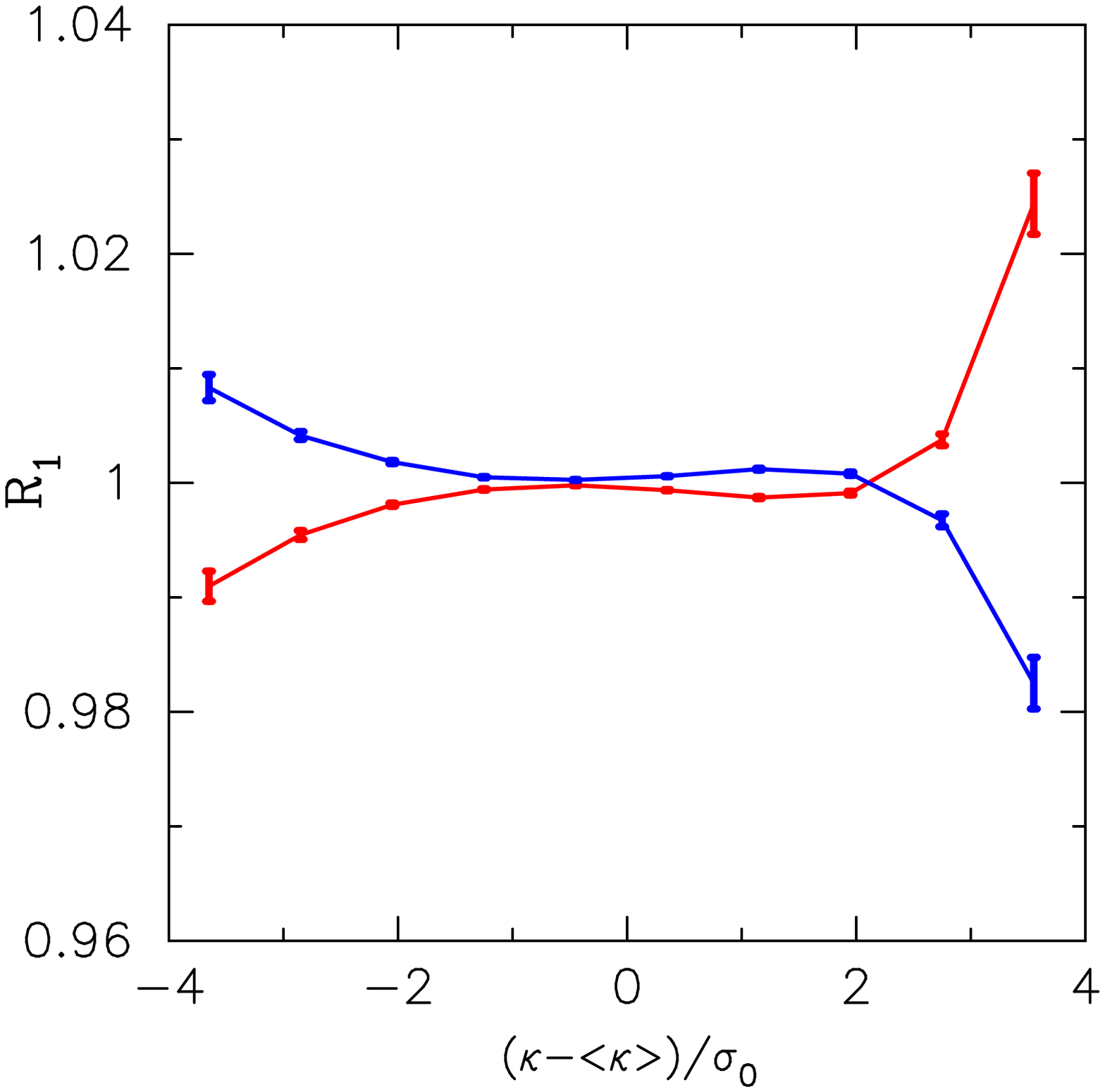}
        \includegraphics[clip, width=0.3\columnwidth]{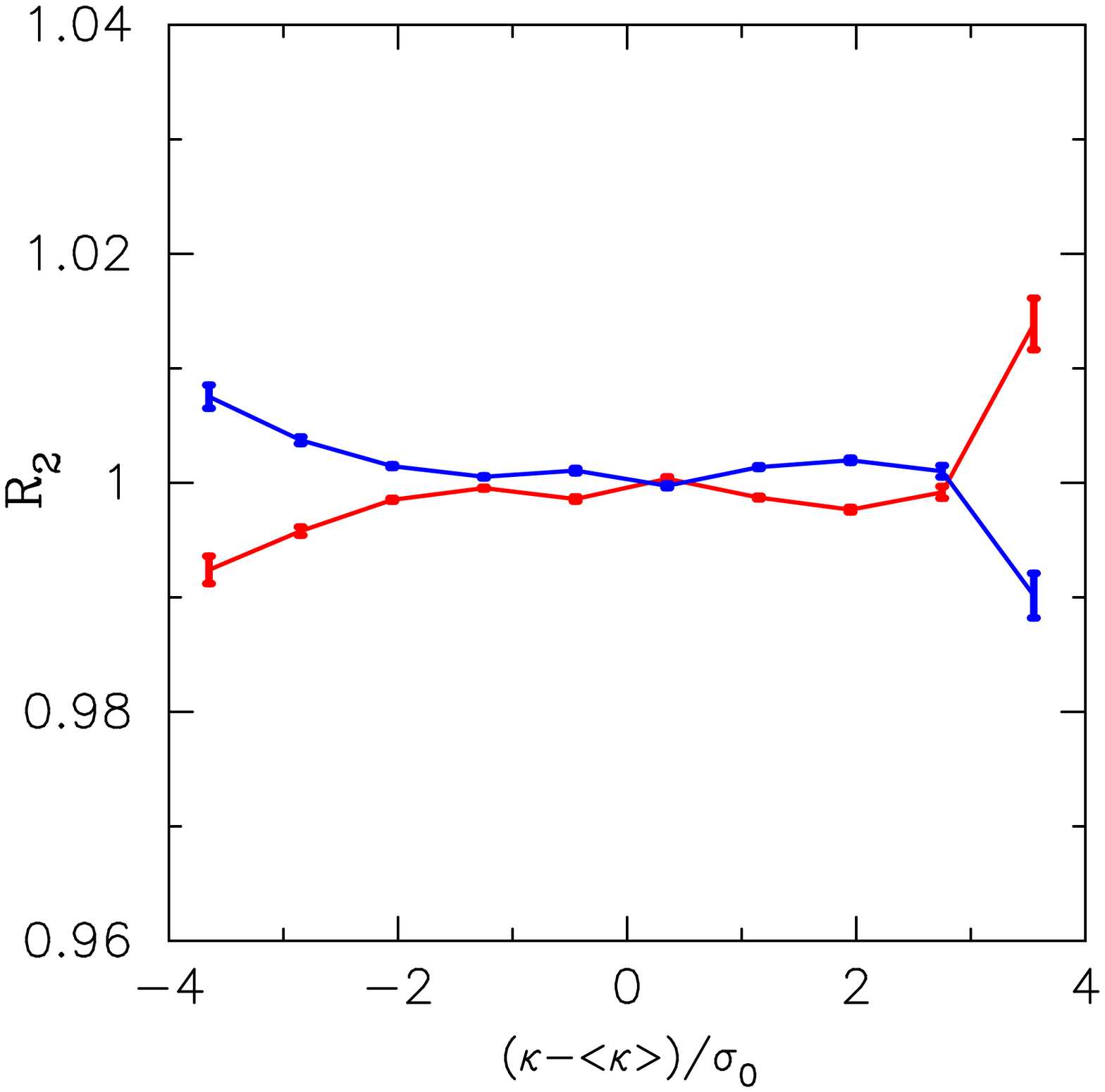}
    	\caption{
	The MFs for our simulated maps with $f_{\rm NL}$.
	The vertical axis shows the ratio of lensing MFs $V_{i}(f_{\rm NL}=\pm 100)/V_{i}(f_{\rm NL}=0)$.
	The horizontal axis shows the normalized convergence field $(\kappa-\langle \kappa \rangle)/\sigma_{0}$.
	The red (blue) points with error bar represent the result from 100 noisy convergence maps with $f_{\rm NL} = +100$ (-100).
	The source plane locates on $z_{\rm source} = 1.0$ and 1 arcmin Gaussian smoothing is adopted.
	 \label{fig:MF_fnl}
	}
    \end{center}
  \end{figure}

Let us first discuss how the primordial non-Gaussianity $f_{\rm NL}$
affects the lensing convergence and the MFs. 
We define the ratio of MFs with respect to
the fiducial $f_{\rm NL} = 0$ model as follows:
\beqa
R_{i} (f_{\rm NL}) = \frac{V_{i}(x \, ; \, f_{\rm NL} \neq 0)}{V_{i}(x \, ; \, f_{\rm NL} = 0)}, \, i = 0,1,2.
\eeqa
Figure \ref{fig:MF_fnl} shows $R_{i}$ from our 100 
convergence maps with $z_{\rm source} =1.0$.
The effect of $f_{\rm NL}$ appears large in the regime where 
the normalized convergence $x$ \psim $3$ for all $V_{i}$s.
This simply reflects the fact 
that $f_{\rm NL}$ affects the number of very massive halos
with mass $\sim 10^{15} h^{-1} M_{\odot}$ which yield $x \psim 3$ 
\citep[see also][]{2004MNRAS.350..893H}.
The abundance of the massive halos at $z=0.5$ is larger 
by $\sim O(10\%)$ for $f_{\rm NL} = 100$
compared to $f_{\rm NL} = 0$ \citep{2010JCAP...07..002N}.
Then the fraction of area  ($V_{0}$) with very high convergence
increases for positive $f_{\rm NL}$, and the total length of contours 
($V_{1}$) increases too.
Interestingly, $f_{\rm NL}$ also affects $V_{1}$ and $V_{2}$ 
at small $x$. 
Because the MFs, $V_{0}, V_{1}, V_{2}$ are {\it not} independent statistics, 
their correlations need to be considered.
We use all the MFs combined together in our statistical analysis below,
in order to extract the full cosmological information and 
to derive an accurate constraint on $f_{\rm NL}$.

\section{RESULT}
\label{sec:res}
\subsection{Fisher analysis}

We perform a Fisher analysis to make forecasts for parameter 
constraints on $f_{\rm NL}$, $w$, and $\sigma_{8}$
for future weak lensing surveys.

For a multivariate Gaussian likelihood, the Fisher matrix $F_{ij}$ 
can be written as
\beqa
F_{ij} = \frac{1}{2} {\rm Tr} \left[ A_{i} A_{j} + C^{-1} M_{ij} \right], \label{eq:Fij}
\eeqa
where $A_{i} = C^{-1} \partial C/\partial \xi_{i}$, 
$M_{ij} = 2 \left(\partial \mu/\partial \xi_{i} \right)\left(\partial \mu/\partial \xi_{j} \right)$, 
$C$ is the data covariance matrix, 
$\mu$ is the assumed model, 
and $\bd{\xi} = (f_{\rm NL},\, w,\, \sigma_{8})$.
For lensing MFs, $\mu$ corresponds to $V_{0}$, $V_{1}$, and $V_{2}$ for different bins.
\footnote{We only consider the second term in Eq.~(\ref{eq:Fij}).
Since $C$ is expected to scale inverse-proportionally to the survey area, 
the second term will be dominant 
for a large area survey \citep{2009A&A...502..721E}.}


We estimate $\mu$ by averaging MFs over our 100 (noisy) convergence maps.
To calculate $M_{ij}$, we approximate the first derivative of MFs 
by the cosmological parameter $\xi_{i}$ as follows
\beqa
\frac{\partial \mu}{\partial \xi_{i}} = \frac{\mu(\xi_{i}+\delta \xi^{(1)}_{i}) - \mu(\xi_{i} + \delta \xi^{(2)}_{i})}{\delta \xi^{(1)}_{i}-\delta \xi^{(2)}_{i}}. \label{eq:dev_MFs}
\eeqa
We use the data set of the three MFs for $z_{\rm source} = 1.0$ or/and $1.5$.
We use 10 bins in the range of $x = [-3, 3]$.
\footnote{In principle, one could use regions with $x>3$, which are
thought to be sensitive to $f_{\rm NL}$. However, such regions are extremely
rare, and thus estimates for the first derivatives in Eq.~ (\ref{eq:dev_MFs})
become uncertain even with our large number of convergence maps.}
We have checked this binning is sufficient to produce robust results
in the following analysis.
In this range of $x$, Eq.~(\ref{eq:dev_MFs}) gives smooth estimate for $M_{ij}$.
In total, we need $60 \times 60$ MFs covariance matrix for the Fisher analysis.
For this purpose, we use 1000 convergence maps made by \citet{2009ApJ...701..945S}.
These maps have the same design as our convergence maps, 
but are generated for slightly different cosmological parameters 
(consistent with WMAP 3-years results \citep{2007ApJS..170..377S}).
We essentially assume that the dependence of the covariance matrix
to cosmological parameters is unimportant. 

We also take into account the constraints from the CMB priors 
expected from the Planck satellite mission.
When we compute the Fisher matrix for the CMB, we use 
the Markov-Chain Monte-Carlo (MCMC) engine for exploring cosmological 
parameter space {\tt COSMOMC} \citep{Lewis:2002ah}.
We consider the parameter constraints from the angular 
power spectra of temperature anisotropies, $E$-mode polarization and 
their cross-correlation.
For MCMC, in addition to $\sigma_{8}$ and $w$, our independent variables 
include the matter density $\Omega_{m} h^2$, 
the baryon density $\Omega_{b} h^2$, 
Hubble parameter $h$, 
reionization optical depth $\tau$, 
and the scalar spectral index $n_s$.
To examine the pure power of lensing MFs to constrain $f_{\rm NL}$, 
we do not include any constraints on $f_{\rm NL}$ from the CMB.
Assuming that the constraints from the CMB and the lensing MFs are independent, 
we express the total Fisher matrix as
\beqa
\bd{F} = \bd{F}_{\rm MFs} + \bd{F}_{\rm CMB}. \label{eq:Ftot}
\eeqa
When we include the CMB priors by Eq.~(\ref{eq:Ftot}), 
we marginalize over the other cosmological parameters 
except $f_{\rm NL}$, $w$ and $\sigma_{8}$.

Strictly speaking, one needs to consider a multivariate non-Gaussian
likelihood because MFs are non-Gaussian estimator.
In the present paper, we employ the Fisher analysis that assumes 
a local Gaussian likelihood in the parameter space. 
Note however that, because
we use fully nonlinear covariance matrices evaluated from 1000 
ray-tracing simulations,
our analysis appropriately includes non-Gaussian error contributions.
Non-Gaussian error will include the contribution of four-point statistics at least
\citep[cf.][]{2012MNRAS.419..536M}.
It is illustrative to show the impact of non-Gaussian errors of MFs 
for parameter estimation.
For comparison, we generate Gaussian covariance matrices
by using 1000 Gaussian convergence maps.
We have found that the resulting constraint on cosmological 
parameters is degraded by a factor of a few percent compared
to the case with Gaussian errors.

\subsection{Forecasts}

We show the forecast for the upcoming survey 
such as Subaru Hyper Suprime-Cam (HSC) and the Large Synoptic Survey Telescope (LSST).
We consider two surveys with an area coverage of 1500 ${\rm deg}^2$ 
and 20000 ${\rm deg}^2$; the former is for HSC, and the latter is for LSST.
We first derive constraints on the cosmological 
parameters for a 25 ${\rm deg}^2$ area survey, 
for which we have the full covariance matrix.
Then we simply scale the covariance matrix by a factor 
of $25/1500 = 1/60$ or $25/20000 = 1/800$ for the two surveys considered.

\begin{figure}
\begin{center}
       \includegraphics[clip, width=0.45\columnwidth]{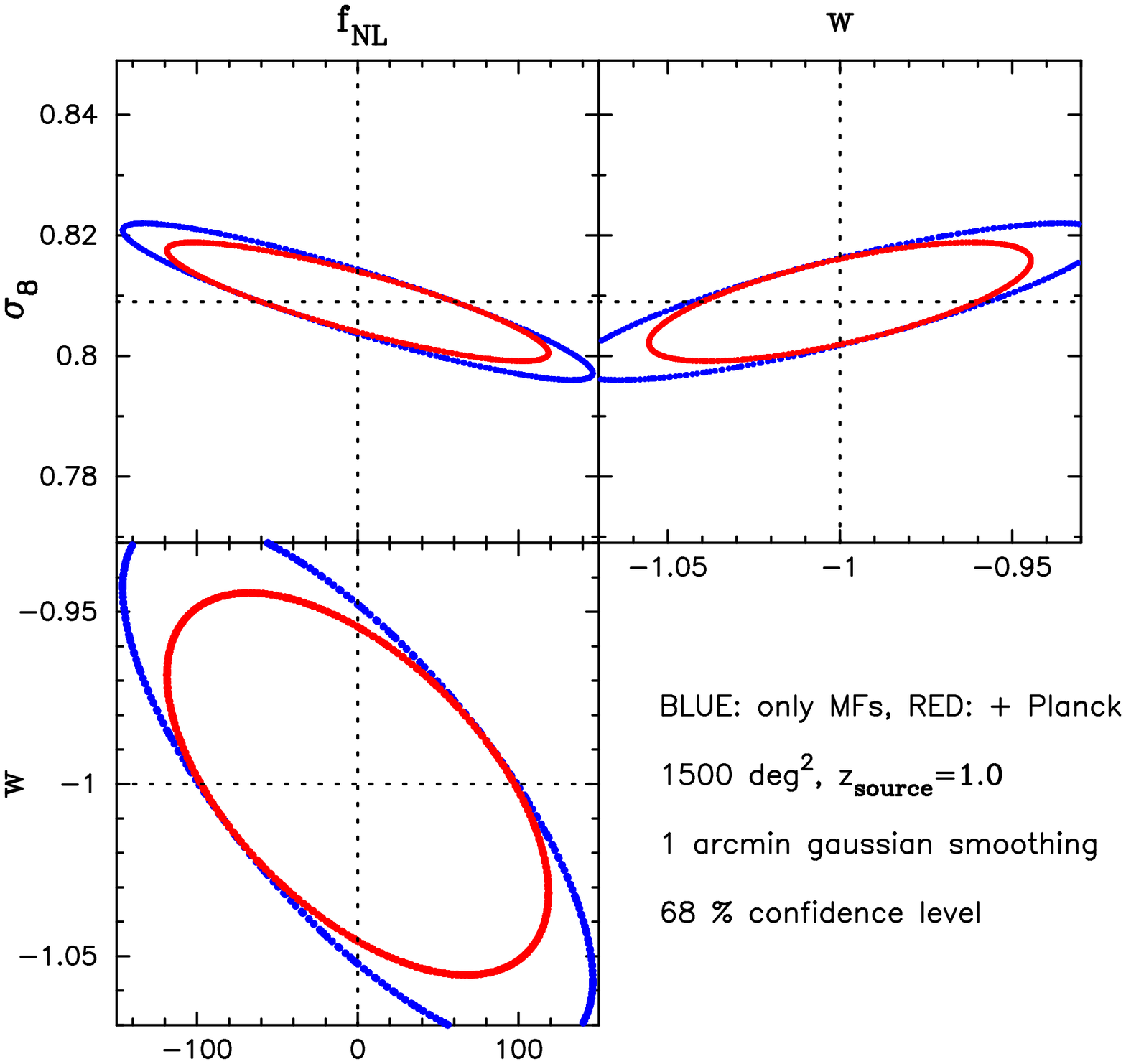}
      \includegraphics[clip, width=0.45\columnwidth]{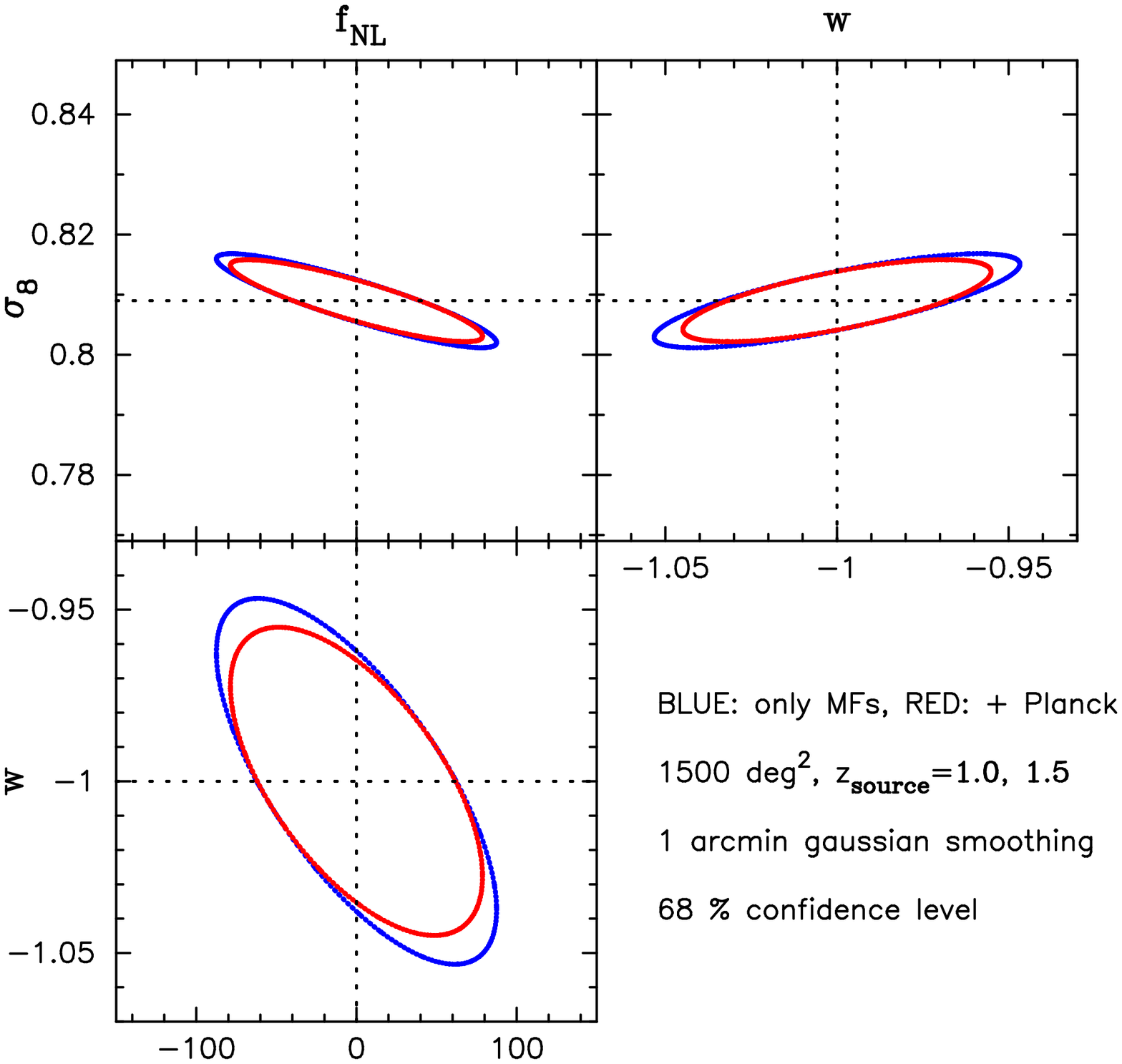}
    \caption{
    We plot 1-$\sigma$ confidence level by weak lensing MFs 
    for Subaru HSC survey (1500 ${\rm deg}^2$).
    The left panel shows the constraints on $f_{\rm NL}$, 
    $w$ and $\sigma_{8}$ for the case with $z_{\rm source} = 1.0$.
    The right panel shows the constraints obtained by a tomographic survey
    with $z_{\rm source} = 1.0$ and $1.5$.
    The blue line shows the constraints from lensing MFs only
    and the red one represents those by MFs and the Planck priors.
    \label{fig:contour_HSC}
    }
    
    \end{center}
\end{figure}

Figure \ref{fig:contour_HSC} shows the two-dimensional confidence 
contours for HSC (1500 ${\rm deg}^2$), in each case marginalized 
over other parameters.
In each panel, the blue line shows the constraint from 
lensing MFs only and the red one is for the case of lensing MFs and 
the Planck priors.
The ellipses shown in this figure correspond to the $68 \%$ 
confidence level from the Fisher analysis.
Our fiducial model in this analysis is 
$(f_{\rm NL}\,, w\,, \sigma_{8}) = (0\,,-1\,,0.809)$.
With the Planck priors, we can constrain $f_{\rm NL}$ 
with a level of $\sim 80$ by using the source plane at $z_{\rm source}=1.0$,
after marginalized over $w$ and $\sigma_{8}$.

For the upcoming multiple-band imaging surveys such as HSC and LSST, 
it is possible to obtain photometric redshifts for the source galaxies.
It is interesting to study how the parameter constraints can be
improved by using source galaxies at higher redshifts.
To this end, we perform the same analysis assuming the source galaxies 
are located at two redshifts, $z_{\rm source}=1.0$ and 1.5.
We show the Fisher analysis result 
in the right panel of Figure \ref{fig:contour_HSC}.
We see the constraint on $f_{\rm NL}$ is improved by a factor of $\sim$ 1.5 
with the two redshift bins.
We summarize the marginalized constraints on $f_{\rm NL}$ in Table \ref{tab:sig_fnl}.

\section{CONCLUSION AND DISCUSSION}
\label{sec:conc}

We have studied the ability of weak lensing MFs
to constrain the local type primordial 
non-Gaussianity.
We have performed 20 $N$-body simulations  
and 100 independent ray-tracing simulations for each
of the seven cosmological models that differ in
$f_{\rm NL}$, $w$ and $\sigma_{8}$. 
We have then performed a Fisher analysis
using the large set of mock lensing maps, 
to obtain confidence limits for cosmological parameters.

The MFs are sensitive probes of $f_{\rm NL}$, 
especially at high convergence values,
because such high convergence regions are associated with
massive halos with \psim $10^{15} h^{-1}M_{\odot}$, 
of which the abundance is sensitive to $f_{\rm NL}$.
We also find that the three MFs, $V_{0},V_{1}$ and $V_{2}$,
are affected differently by $f_{\rm NL}$. 
This means that combining the three MFs gives 
tighter constraints on $f_{\rm NL}$.

From a Fisher analysis, we have obtained the following results.
For source galaxies at $z_{\rm source} = 1.0$,  
the primordial non-Gaussianity is constrained with a level of $f_{\rm NL} \sim 80$ 
for HSC survey with a 1500 ${\rm deg}^2$ survey area.
The constraints can be improved by selecting source galaxies
in multiple redshift bins. 
We find that the constraint on $f_{\rm NL}$ is improved to $\sim 50$,
i.e., by a factor of $\sim$ 1.5 if we use source galaxies 
at $z_{\rm source} = 1.0$ and $1.5$.
This is largely because the degeneracy between $\sigma_8$ 
and $w$ is broken due to information contained in the
matter distribution at different redshifts.
We have also tested how the number density of the source galaxies 
affects our analysis.
We have re-analyzed the case of 
$n_{\rm gal}=30$ galaxies$/{\rm arcmin}^2$ for a fixed source plane
at $z_{\rm source}=1.0$.
The constraint on $f_{\rm NL}$ is improved only by a factor 
of $\sim$ 5 \% in this case.
We argue that the ``tomographic'' information using the multiple 
source planes is useful to derive accurate constraints on $f_{\rm NL}$, 
even though one then needs to use a smaller number of source galaxies 
at each of the source planes.

Ultimately, for a LSST-like survey with a 20000 ${\rm deg}^2$ area,
we can obtain the constraint of $f_{\rm NL} \sim 25$ with $z_{\rm source} = 1.0$, 
and $f_{\rm NL} \sim 15$ with $z_{\rm source} = 1.0$ and $1.5$.
In principle, these constraints will be further improved by 
including the high sigma bins of MFs because the higher convergence 
region is more sensitive to $f_{\rm NL}$.
We will continue our study along this line using a larger
set of simulations of larger volumes.

Finally, we discuss possible technical improvements
in using the MFs of weak lensing maps. 
First, we have checked how the smoothing scale on the convergence maps 
affects our analysis. We have performed a Fisher analysis
using the maps with smoothing of 0.5 arcmin, 1 arcmin, 2 arcmin and 5 arcmin.
Smoothing with $\sim 1$ arcmin has turned out to be optimal
for the targeted future surveys, yielding the best constraints.
This is explained qualitatively by the fact that 
only the large-scale, linear structure is probed with large smoothing scales,
whereas for smaller smoothing scales, the intrinsic shape noise becomes large. 
We note that the angular-scale dependence of MFs itself can provide more information.
For example, the angular-scale dependence can be used to separate primordial 
non-Gaussianities from gravity-induced non-Gaussianities
\citep[e.g.][]{2012MNRAS.419..138M, 2012MNRAS.419..536M, 2012PhRvD..85j3513K}.
One could further improve the cosmological constraints by combining 
MFs with various smoothing scales and their evolutions.
Evaluating MFs is complicated when an observed map include masked regions.
Also, instrumental and atmospheric systematics can easily compromise the 
measurement of lensing MFs. These issues certainly warrants further extensive studies.
The upcoming wide-field surveys will provide highly-resolved lensing maps.
Our study in the present paper may be useful to properly analyze the data
and extract cosmological information from them.

\acknowledgments

We thank Chiaki Hikage and Masamune Oguri for useful discussions.
Masanori Sato provided us with their ray-tracing simulations data.
M.S. is supported by Research Fellowships of the Japan Society for 
the Promotion of Science (JSPS) for Young Scientists.
T.N. is supported by a Grant-in-Aid for the JSPS fellows. 
This work is supported by World Premier International Research Center 
Initiative (WPI Initiative), MEXT, Japan
and in part by Grant-in-Aid for Scientific Research from the JSPS Promotion of Science (23540324).
Numerical computations presented in this paper were in part carried out
on the general-purpose PC farm at Center for Computational Astrophysics,
CfCA, of National Astronomical Observatory of Japan.

\bibliography{ref}

\begin{thebibliography}{35}
\expandafter\ifx\csname natexlab\endcsname\relax\def\natexlab#1{#1}\fi

\bibitem[{{Bartolo} {et~al.}(2004){Bartolo}, {Komatsu}, {Matarrese}, \&
  {Riotto}}]{2004PhR...402..103B}
{Bartolo}, N., {Komatsu}, E., {Matarrese}, S., \& {Riotto}, A. 2004, \physrep,
  402, 103

\bibitem[{{Crocce} {et~al.}(2006){Crocce}, {Pueblas}, \&
  {Scoccimarro}}]{2006MNRAS.373..369C}
{Crocce}, M., {Pueblas}, S., \& {Scoccimarro}, R. 2006, \mnras, 373, 369

\bibitem[{{Croton} {et~al.}(2007){Croton}, {Gao}, \&
  {White}}]{2007MNRAS.374.1303C}
{Croton}, D.~J., {Gao}, L., \& {White}, S.~D.~M. 2007, \mnras, 374, 1303

\bibitem[{{Dalal} {et~al.}(2008){Dalal}, {Dor{\'e}}, {Huterer}, \&
  {Shirokov}}]{2008PhRvD..77l3514D}
{Dalal}, N., {Dor{\'e}}, O., {Huterer}, D., \& {Shirokov}, A. 2008, \prd, 77,
  123514

\bibitem[{{de Bernardis} {et~al.}(2010){de Bernardis}, {Serra}, {Cooray}, \&
  {Melchiorri}}]{2010PhRvD..82h3511D}
{de Bernardis}, F., {Serra}, P., {Cooray}, A., \& {Melchiorri}, A. 2010, \prd,
  82, 083511

\bibitem[{{Eifler} {et~al.}(2009){Eifler}, {Schneider}, \&
  {Hartlap}}]{2009A&A...502..721E}
{Eifler}, T., {Schneider}, P., \& {Hartlap}, J. 2009, \aap, 502, 721

\bibitem[{{Hamana} \& {Mellier}(2001)}]{2001MNRAS.327..169H}
{Hamana}, T., \& {Mellier}, Y. 2001, \mnras, 327, 169

\bibitem[{{Hamana} {et~al.}(2004){Hamana}, {Takada}, \&
  {Yoshida}}]{2004MNRAS.350..893H}
{Hamana}, T., {Takada}, M., \& {Yoshida}, N. 2004, \mnras, 350, 893

\bibitem[{{Hikage} {et~al.}(2008){Hikage}, {Matsubara}, {Coles}, {Liguori},
  {Hansen}, \& {Matarrese}}]{2008MNRAS.389.1439H}
{Hikage}, C., {Matsubara}, T., {Coles}, P., {et~al.} 2008, \mnras, 389, 1439

\bibitem[{{Komatsu} \& {Spergel}(2001)}]{2001PhRvD..63f3002K}
{Komatsu}, E., \& {Spergel}, D.~N. 2001, \prd, 63, 063002

\bibitem[{{Komatsu} {et~al.}(2011){Komatsu}, {Smith}, {Dunkley}, {Bennett},
  {Gold}, {Hinshaw}, {Jarosik}, {Larson}, {Nolta}, {Page}, {Spergel},
  {Halpern}, {Hill}, {Kogut}, {Limon}, {Meyer}, {Odegard}, {Tucker}, {Weiland},
  {Wollack}, \& {Wright}}]{2011ApJS..192...18K}
{Komatsu}, E., {Smith}, K.~M., {Dunkley}, J., {et~al.} 2011, \apjs, 192, 18

\bibitem[{{Kratochvil} {et~al.}(2012){Kratochvil}, {Lim}, {Wang}, {Haiman},
  {May}, \& {Huffenberger}}]{2012PhRvD..85j3513K}
{Kratochvil}, J.~M., {Lim}, E.~A., {Wang}, S., {et~al.} 2012, \prd, 85, 103513

\bibitem[{Lewis \& Bridle(2002)}]{Lewis:2002ah}
Lewis, A., \& Bridle, S. 2002, Phys. Rev., D66, 103511

\bibitem[{Lewis {et~al.}(2000)Lewis, Challinor, \& Lasenby}]{Lewis:1999bs}
Lewis, A., Challinor, A., \& Lasenby, A. 2000, Astrophys. J., 538, 473

\bibitem[{{Lim} \& {Simon}(2012)}]{2012JCAP...01..048L}
{Lim}, E.~A., \& {Simon}, D. 2012, \jcap, 1, 48

\bibitem[{{Lo Verde} {et~al.}(2008){Lo Verde}, {Miller}, {Shandera}, \&
  {Verde}}]{2008JCAP...04..014L}
{Lo Verde}, M., {Miller}, A., {Shandera}, S., \& {Verde}, L. 2008, \jcap, 4, 14

\bibitem[{{Matsubara} \& {Jain}(2001)}]{2001ApJ...552L..89M}
{Matsubara}, T., \& {Jain}, B. 2001, \apjl, 552, L89

\bibitem[{{Miyazaki} {et~al.}(2007){Miyazaki}, {Hamana}, {Ellis}, {Kashikawa},
  {Massey}, {Taylor}, \& {Refregier}}]{2007ApJ...669..714M}
{Miyazaki}, S., {Hamana}, T., {Ellis}, R.~S., {et~al.} 2007, \apj, 669, 714

\bibitem[{{Munshi} {et~al.}(2012{\natexlab{a}}){Munshi}, {Smidt}, {Joudaki}, \&
  {Coles}}]{2012MNRAS.419..138M}
{Munshi}, D., {Smidt}, J., {Joudaki}, S., \& {Coles}, P. 2012{\natexlab{a}},
  \mnras, 419, 138

\bibitem[{{Munshi} {et~al.}(2012{\natexlab{b}}){Munshi}, {van Waerbeke},
  {Smidt}, \& {Coles}}]{2012MNRAS.419..536M}
{Munshi}, D., {van Waerbeke}, L., {Smidt}, J., \& {Coles}, P.
  2012{\natexlab{b}}, \mnras, 419, 536

\bibitem[{{Nishimichi} {et~al.}(2010){Nishimichi}, {Taruya}, {Koyama}, \&
  {Sabiu}}]{2010JCAP...07..002N}
{Nishimichi}, T., {Taruya}, A., {Koyama}, K., \& {Sabiu}, C. 2010, \jcap, 7, 2

\bibitem[{{Nishimichi} {et~al.}(2009){Nishimichi}, {Shirata}, {Taruya},
  {Yahata}, {Saito}, {Suto}, {Takahashi}, {Yoshida}, {Matsubara}, {Sugiyama},
  {Kayo}, {Jing}, \& {Yoshikawa}}]{2009PASJ...61..321N}
{Nishimichi}, T., {Shirata}, A., {Taruya}, A., {et~al.} 2009, \pasj, 61, 321

\bibitem[{{Oguri} {et~al.}(2012){Oguri}, {Bayliss}, {Dahle}, {Sharon},
  {Gladders}, {Natarajan}, {Hennawi}, \& {Koester}}]{2012MNRAS.420.3213O}
{Oguri}, M., {Bayliss}, M.~B., {Dahle}, H., {et~al.} 2012, \mnras, 420, 3213

\bibitem[{{Oguri} \& {Takada}(2011)}]{2011PhRvD..83b3008O}
{Oguri}, M., \& {Takada}, M. 2011, \prd, 83, 023008

\bibitem[{{Reid} {et~al.}(2010{\natexlab{a}}){Reid}, {Verde}, {Dolag},
  {Matarrese}, \& {Moscardini}}]{2010JCAP...07..013R}
{Reid}, B.~A., {Verde}, L., {Dolag}, K., {Matarrese}, S., \& {Moscardini}, L.
  2010{\natexlab{a}}, \jcap, 7, 13

\bibitem[{{Reid} {et~al.}(2010{\natexlab{b}}){Reid}, {Percival}, {Eisenstein},
  {Verde}, {Spergel}, {Skibba}, {Bahcall}, {Budavari}, {Frieman}, {Fukugita},
  {Gott}, {Gunn}, {Ivezi{\'c}}, {Knapp}, {Kron}, {Lupton}, {McKay}, {Meiksin},
  {Nichol}, {Pope}, {Schlegel}, {Schneider}, {Stoughton}, {Strauss}, {Szalay},
  {Tegmark}, {Vogeley}, {Weinberg}, {York}, \& {Zehavi}}]{2010MNRAS.404...60R}
{Reid}, B.~A., {Percival}, W.~J., {Eisenstein}, D.~J., {et~al.}
  2010{\natexlab{b}}, \mnras, 404, 60

\bibitem[{{Sato} {et~al.}(2001){Sato}, {Takada}, {Jing}, \&
  {Futamase}}]{2001ApJ...551L...5S}
{Sato}, J., {Takada}, M., {Jing}, Y.~P., \& {Futamase}, T. 2001, \apjl, 551, L5

\bibitem[{{Sato} {et~al.}(2009){Sato}, {Hamana}, {Takahashi}, {Takada},
  {Yoshida}, {Matsubara}, \& {Sugiyama}}]{2009ApJ...701..945S}
{Sato}, M., {Hamana}, T., {Takahashi}, R., {et~al.} 2009, \apj, 701, 945

\bibitem[{{Shandera} {et~al.}(2011){Shandera}, {Dalal}, \&
  {Huterer}}]{2011JCAP...03..017S}
{Shandera}, S., {Dalal}, N., \& {Huterer}, D. 2011, \jcap, 3, 17

\bibitem[{{Spergel} {et~al.}(2007){Spergel}, {Bean}, {Dor{\'e}}, {Nolta},
  {Bennett}, {Dunkley}, {Hinshaw}, {Jarosik}, {Komatsu}, {Page}, {Peiris},
  {Verde}, {Halpern}, {Hill}, {Kogut}, {Limon}, {Meyer}, {Odegard}, {Tucker},
  {Weiland}, {Wollack}, \& {Wright}}]{2007ApJS..170..377S}
{Spergel}, D.~N., {Bean}, R., {Dor{\'e}}, O., {et~al.} 2007, \apjs, 170, 377

\bibitem[{{Springel}(2005)}]{2005MNRAS.364.1105S}
{Springel}, V. 2005, \mnras, 364, 1105

\bibitem[{{Tegmark} {et~al.}(2006){Tegmark}, {Eisenstein}, {Strauss},
  {Weinberg}, {Blanton}, {Frieman}, {Fukugita}, {Gunn}, {Hamilton}, {Knapp},
  {Nichol}, {Ostriker}, {Padmanabhan}, {Percival}, {Schlegel}, {Schneider},
  {Scoccimarro}, {Seljak}, {Seo}, {Swanson}, {Szalay}, {Vogeley}, {Yoo},
  {Zehavi}, {Abazajian}, {Anderson}, {Annis}, {Bahcall}, {Bassett}, {Berlind},
  {Brinkmann}, {Budavari}, {Castander}, {Connolly}, {Csabai}, {Doi},
  {Finkbeiner}, {Gillespie}, {Glazebrook}, {Hennessy}, {Hogg}, {Ivezi{\'c}},
  {Jain}, {Johnston}, {Kent}, {Lamb}, {Lee}, {Lin}, {Loveday}, {Lupton},
  {Munn}, {Pan}, {Park}, {Peoples}, {Pier}, {Pope}, {Richmond}, {Rockosi},
  {Scranton}, {Sheth}, {Stebbins}, {Stoughton}, {Szapudi}, {Tucker}, {vanden
  Berk}, {Yanny}, \& {York}}]{2006PhRvD..74l3507T}
{Tegmark}, M., {Eisenstein}, D.~J., {Strauss}, M.~A., {et~al.} 2006, \prd, 74,
  123507

\bibitem[{{Tomita}(1986)}]{1986PThPh..76..952T}
{Tomita}, H. 1986, Progress of Theoretical Physics, 76, 952

\bibitem[{{Valageas} \& {Nishimichi}(2011)}]{2011A&A...527A..87V}
{Valageas}, P., \& {Nishimichi}, T. 2011, \aap, 527, A87

\bibitem[{{White} \& {Hu}(2000)}]{2000ApJ...537....1W}
{White}, M., \& {Hu}, W. 2000, \apj, 537, 1

\end{thebibliography}

\clearpage

\begin{table}
\begin{center}
\begin{tabular}{|c|c|c|c|c|c|c|}
\tableline
& $f_{\rm NL}$ & $w_{\rm DE}$ & $\sigma_{8}$ & \# of $N$-body sims & \# of maps\\ \tableline
fiducial & 0 & -1.0 & 0.809& 20 & 100\\ \tableline
high $w_{\rm DE}$ & 0 & -0.8 & 0.753& 20 & 100\\ \tableline
low $w_{\rm DE}$ & 0 & -1.2 & 0.848& 20 & 100\\ \tableline
high $f_{\rm NL}$ & 100 & -1.0 & 0.809& 20 & 100\\ \tableline
low $f_{\rm NL}$ & -100 & -1.0 & 0.809& 20 & 100\\ \tableline
high $\sigma_{8}$ & 0 & -1.0 & 0.848& 20 & 100\\ \tableline
low $\sigma_{8}$ & 0 & -1.0 & 0.753& 20 & 100\\ \tableline
\end{tabular} 
\caption{Parameters for our $N$-body simulations.
	For each model, we run 20 $N$-body 
        realizations and generate 100 weak lensing convergence maps.}
\label{tab:nbody}
\end{center}
\end{table}

\begin{table}
\begin{center}
\begin{tabular}{|c|c|c|}
\tableline
& $z_{\rm source} = 1.0$ & $z_{\rm source} = 1.0,\, 1.5$ \\ \tableline
MFs only (1500 ${\rm deg}^2$) & 96.8 &  57.9 \\ \tableline
MFs + Planck (1500 ${\rm deg}^2$) & 78.5 & 52.1 \\ \tableline
MFs only (20000 ${\rm deg}^2$) & 26.5 &  15.8 \\ \tableline
MFs + Planck (20000 ${\rm deg}^2$) & 25.8 & 15.7 \\ \tableline
\end{tabular} 
\caption{
  The 1-$\sigma$ constraint on $f_{\rm NL}$ when marginalized over $w$ and $\sigma_{8}$.
  We consider two surveys with a survey area of 1500 ${\rm deg}^2$ (HSC)
  and 20000 ${\rm deg}^2$ (LSST).
  The analysis includes the intrinsic noise from source galaxies 
  with the number density of $n_{\rm gal}=15 \ {\rm galaxies}/{\rm arcmin}^{2}$ 
  at each source redshift.
  \label{tab:sig_fnl}
}
\end{center}
\end{table}

\end{document}